\documentclass[sigplan,screen,nonacm]{acmart}
\usepackage{xcolor}
\usepackage{algorithm}
\usepackage{algpseudocode}
\AtBeginDocument{%
  }

\setcopyright{none}
\makeatletter
\def\@mkauthors@iii{%
  \global\setbox\mktitle@bx=\vbox{%
    \unvbox\mktitle@bx\par\medskip
    \centering\normalfont\normalsize
    Ruifeng Zhang\textsuperscript{1}\quad
    Sai Krishna Teja Varma Manthena\textsuperscript{1}\quad
    Jiajia Li\textsuperscript{1}\quad
    Xipeng Shen\textsuperscript{2}\par
    \vspace{2pt}\small
    \textsuperscript{1}North Carolina State University, USA\qquad
    \textsuperscript{2}Purdue University, USA\par
    \vspace{2pt}\footnotesize
    \{rzhang38, smanthe, jiajia.li\}@ncsu.edu\qquad shen810@purdue.edu\par
    \medskip}}
\makeatother

\begin{document}

\title{Scaling Fourier-Based Sparse Matrix Analysis on GPUs}

\author{Ruifeng Zhang}
\affiliation{\institution{North Carolina State University}\country{USA}}
\email{rzhang38@ncsu.edu}

\author{Sai Krishna Teja Varma Manthena}
\affiliation{\institution{North Carolina State University}\country{USA}}
\email{smanthe@ncsu.edu}

\author{Jiajia Li}
\affiliation{\institution{North Carolina State University}\country{USA}}
\email{jiajia.li@ncsu.edu}

\author{Xipeng Shen}
\affiliation{\institution{Purdue University}\country{USA}}
\email{shen810@purdue.edu}

\renewcommand{\shortauthors}{Zhang et al.}

\begin{abstract}
Sparse computations are important workloads in applications such as scientific computing, graph neural networks (GNNs), and machine learning. While many sparse operations can benefit from modern GPUs, the sparsity pattern remains important to performance because it affects memory coalescing, block organization, and load balancing. Previous studies show that spectral signatures can help analyze the global structure of sparse matrices. The fast Fourier transform (FFT) is commonly used to extract spectral signatures, and efficient GPU FFT libraries are available. However, sparse matrices, especially adjacency matrices for large graphs, tend to be very large and sparse. Existing dense-matrix-based FFT implementations are difficult to scale up, making the spectral patterns of these matrices difficult to obtain. We therefore propose a three-fold research approach comprising a lossless Binary-Sparse FFT (BS-FFT) and two compression methods: \textcolor{black}{Elastic BS-FFT, which reuses} the BS-FFT pipeline on a sampled frequency grid, and \textcolor{black}{density-map-based spatial compression}. Experiments show that \textcolor{black}{BS-FFT reduces GPU memory use by 2.9--11.6$\times$ relative to dense cuFFT and completes} all 15 GNN adjacency matrices where dense cuFFT completes 6 on a 40~\textcolor{black}{GB} A100. \textcolor{black}{Elastic BS-FFT and \textcolor{black}{Density Map compression} reduce GPU computation time by 2.0--1466.4$\times$ relative to BS-FFT with spectral feature error of only 0.16\% to 11.56\% across the sampling rates from 6.25\% to 0.0061\%.}
\end{abstract}



\maketitle
\pagestyle{plain}
\thispagestyle{plain}

\section{Introduction}

Sparse computations are fundamental to scientific computing~\cite{lu2020efficient,ginkgo-toms-2022}, graph analytics~\cite{wang2019deep,huang2020ge}, and machine learning~\cite{gale2020sputnik,zheng2022sparta}, yet their performance is highly sensitive to the sparsity pattern~\cite{zhou2019enabling,yesil2023wise}. Cache reuse, memory coalescing, tile occupancy, and load balance depend not only on matrix dimensions and nonzero count, but also on where the nonzeros occur. Understanding this structure is therefore important for characterizing sparse workloads and reasoning about sparse-kernel performance.

\begingroup\emergencystretch=2em Spectral analysis offers a global view of that structure. Prior work~\cite{zhang2026spectralanalysissparsematrix} treats a sparse matrix as a two-dimensional signal and analyzes its Fourier representation. \color{black} It demonstrates that these signatures reveal
structural scale, orientation, and regularity that complement
spatial statistics and improve applications such as sparse-kernel selection. On
SuiteSparse~\cite{DavisHu2011} matrices, adding spectral features to the WISE
spatial features increases SpMV format-selection accuracy. On pruned
LLM matrices, the same combination yields up to 24.5\%
selected-kernel speedups over using spatial features
alone. These results establish the practical value of spectral
analysis and motivate making its extraction affordable on
larger sparse matrices. \color{black}\par\endgroup

Computing these signatures on GPUs at realistic scale is difficult. Public graph-learning datasets can contain hundreds of millions of nodes and billions of edges~\cite{Hu2021OGBLSC}, and their adjacency matrices tend to be very large and sparse. Note that spatial sparsity also does not imply frequency sparsity: every input nonzero contributes a phase term to every Fourier coefficient, so a sparse pattern often produces a dense spectrum. Conventional GPU FFT libraries such as cuFFT~\cite{NVIDIAcuFFT} nevertheless materialize the complete dense input and output. At $100k\times100k$, those two dense arrays alone require roughly \textcolor{black}{80~GB} before FFT workspace, exceeding a 40~\textcolor{black}{GB} A100. For an $m\times n$ input, moving the transform to the CPU may provide a larger aggregate memory capacity, but it does not change the fundamental \(O(mn)\) dense storage requirement or the \(O(mn(\log m+\log n))\) computational complexity. In practice, CPU execution is substantially slower and still encounters memory-capacity limits at sufficiently large problem sizes.

Existing techniques solve only parts of this problem. Sparse FFT algorithms aim to recover a few dominant coefficients by assuming a sparse or compressible spectrum~\cite{Hsieh2014SFFTDT,Wang2017FPSSFT}, \textcolor{black}{an assumption that may not hold as the spectrum is often dense}. Some other sparse FFT libraries support the fast calculation of a set of frequency points~\cite{ETHCSCSpFFT,Barnett2019FINUFFT}, but they primarily target settings where only a small number of spectral coefficients are required and do not exploit sparsity in the input domain. Distributed and multi-GPU systems increase aggregate capacity and throughput, but still store dense arrays proportional to matrix area~\cite{Ayala2020heFFTe,NVIDIAcuFFTMp}. To our knowledge, prior work has not studied scaling up FFT on large, sparse inputs.

How to effectively utilize the sparse input is a non-trivial problem. Directly evaluating the whole spectrum from $K$ nonzeros costs $O(mnK)$ because each requested coefficient depends on the complete sparse input. A naive GPU mapping, therefore, scans the same nonzeros repeatedly, returning every coefficient. However, it retains an $O(mn)$ capacity cost regardless of how compactly the input is stored. Therefore, how to compress the output is also a challenge. Meanwhile, many pattern-analysis tasks consume summaries such as radial and angular energy, low-frequency concentration, or spectral entropy~\cite{Cao2019FrequencySpectrum,Inouye1991SpectralEntropy} rather than the complete spectrum, so another challenge is restoring these spectral features from the compressed spectrum.

We therefore explore three methods for different scales and accuracy--cost trade-offs: the Binary-Sparse FFT (BS-FFT), its sampled variant Elastic BS-FFT, and density-map-based spatial compression.

First, \emph{BS-FFT} performs exact full-spectrum computation. It transforms a binary sparse matrix directly from its nonzero indices, never forming the dense grid, and halves both stages by conjugate symmetry. Because it still returns the full dense spectrum, it extends feasibility but remains output-bound on the largest inputs. Therefore, we propose \emph{Elastic BS-FFT}, which evaluates coefficients on a Cartesian subset of the original frequency lattice using the BS-FFT pipeline. It avoids both dense input and output without assuming a sparse spectrum. The retained coefficients are exact, and we demonstrate theoretically and empirically that this sampling method achieves superior accuracy and performance. However, it still incurs substantial overhead at relatively high sampling rates. To this end, we propose \emph{density-map-based spatial compression (Density Map).} The Density Map compression aggregates all nonzeros into a smaller occupancy grid and applies cuFFT to that grid. It is naturally fast because it directly compresses the input and requires minimal optimization effort. However, distinct matrices can therefore produce the same density map while having different spectra, so spatial compression cannot preserve the exact Fourier coefficients.

Together, these paths form a scale ladder: BS-FFT preserves the complete spectrum while capacity permits, Elastic BS-FFT spends additional computation to keep retained coefficients exact under a bounded output, and density maps provide an inexpensive resolution. To our knowledge, this work presents the first scaling ladder
for spectral analysis of large sparse matrices on modern GPUs. To summarize, this paper makes the following contributions: 
\begingroup 
\emergencystretch=2em
\begin{itemize}
  \item The first \textcolor{black}{scale ladder} for scaling up spectral analysis on sparse matrices on modern GPUs.
  \item A GPU sparse FFT implementation (BS-FFT) that computes the exact full spectrum from sparse input: a row transform evaluated directly from the nonzeros, a direct or Bluestein column FFT chosen by the length, a conjugate-symmetric transform that halves both stages, and a double-buffered tile pipeline that bounds working memory. Together they remove the dense spatial grid and extend exact transforms to sizes dense cuFFT cannot plan or fit (Section~\ref{sec:exact-cuspfft}).
  \item An elastic version of BS-FFT which reuses the BS-FFT pipeline but on a sampled uniform Cartesian grid in the spectral domain. \textcolor{black}{We provide an analysis of and insights into why} this sampling is better than heuristic-based energy-directed sampling used by existing sparse libraries (Section~\ref{sec:spectrum-sampling}).
  \item A density-map-based spatial compression (Density Map) approach that compresses sparse input \textcolor{black}{and applies a dense FFT directly}. We further derive \textcolor{black}{a theoretical basis for the sampling-ratio--error trade-off} for \textcolor{black}{Density Map compression} (Section~\ref{sec:density-compression}).
  \item We propose full-spectrum reconstruction via bilinear interpolation and direct spectral feature extraction from the compressed spectrum to support sparse matrix pattern analysis from compressed spectral outputs (\textcolor{black}{\hypersetup{allcolors=black}Sections~\ref{sec:full-spectrum-reconstruction} and~\ref{sec:direct-features}}).
  \item Evaluation on an A100 40~\textcolor{black}{GB} shows that BS-FFT completes all 15 GNN adjacency matrices against 6 for dense cuFFT \textcolor{black}{with GPU memory use reduced by 2.9--11.6$\times$}. Elastic BS-FFT and \textcolor{black}{Density Map compression} reduce GPU computation time by 2.0--1466.4$\times$ relative to BS-FFT with spectral feature error of only 0.16\% to 11.56\% across the sampling rates from 6.25\% to 0.0061\%. \textcolor{black}{(Section~\ref{sec:evaluation}).}
\end{itemize}
\par\endgroup 

\section{Background}

\subsection{2-D Fourier Analysis and Spectral Signatures}

The discrete Fourier transform (DFT) decomposes a spatial signal into complex coefficients indexed by spatial frequency; the fast Fourier transform (FFT) computes the same coefficients more efficiently. For a \textcolor{black}{2-D matrix} $A\in \mathbb{R}^{m\times n}$, the two-dimensional DFT is
\vspace{-0.3\baselineskip}
\begin{equation}
F[u,v] = \sum_{i=0}^{m-1}\sum_{j=0}^{n-1}
A[i,j]e^{-2\pi\mathbf{i}(ui/m+vj/n)},
\label{eq:dft2}
\end{equation}
where $F\in\mathbb{C}^{m\times n}$. The magnitude spectrum is
\vspace{-0.3\baselineskip}
\begin{equation}
M[u,v]=|F[u,v]|
\label{eq:spectral-magnitudes}
\end{equation}
The magnitude records the strength of each spatial-frequency component, whereas the phase records spatial alignment. Normally, an FFT shift is applied before analysis so that the zero-frequency component lies at the center of the spectrum.

Pattern-analysis applications~\cite{Shen1998SpectralEntropy,Oliva2001SpatialEnvelope,Manjunath1996TextureFeatures} often aggregate the $m\times n$\ complete magnitude spectrum into compact descriptors. Let
\vspace{-0.3\baselineskip}
\begin{equation}
p_{u,v}=\frac{M[u,v]^2}{\sum_{\alpha=0}^{m-1}\sum_{\beta=0}^{n-1}M[\alpha,\beta]^2}
\label{eq:normalized-spectral-mass}
\end{equation}
denote normalized spectral power. Radial bins summarize energy by distance from the center, angular sectors summarize directional energy, and $-\sum_{u,v}p_{u,v}\log p_{u,v}$ gives spectral entropy. Low-frequency ratios, smoothness statistics, and directional imbalance provide additional compact views. Prior work~\cite{zhang2026spectralanalysissparsematrix} shows that these descriptors distinguish clustered, banded, periodic, random, and block-structured patterns that can share similar density or row-length statistics.

\subsection{Limitations of Existing Sparse FFT Methods}
\label{sec:existing-fft-methods}

\begingroup\emergencystretch=2em
Existing sparse FFT algorithms target a sparse or compressible frequency domain. sFFT-DT~\cite{Hsieh2014SFFTDT} decodes aliased buckets after downsampling, while FPS-SFT~\cite{Wang2017FPSSFT} uses projection-slice measurements to recover a limited set of dominant coefficients. ETH-CSCS SpFFT~\cite{ETHCSCSpFFT} accepts sparse frequency indices but still transforms a complete space-domain grid. FINUFFT~\cite{Barnett2019FINUFFT} evaluates nonuniform sources at caller-selected targets, but still relies on a dense FFT on an internal oversampled grid. SparseFFT.jl~\cite{HoSparseFFTjl} 
applies a pruned blockwise factorization to requested coordinates, but its 2D implementation retains dense intermediate
buffers that scale with the full transform dimensions. Their native assumptions that only a few dominant spectral coefficients are significant do not generally hold when a sparse input has a dense spectrum, and do not address the scalability challenges of extracting spectral signatures from large sparse matrices.
\par\endgroup

Distributed FFT systems such as heFFTe and cuFFTMp partition dense arrays across GPUs or nodes using slab or pencil decompositions~\cite{Ayala2020heFFTe,NVIDIAcuFFTMp}. Each axis transform requires a global transpose, so parallel FFTs are \textcolor{black}{communication-bound} at scale. In heFFTe's $1024^3$ profile, a $42\times$ acceleration of the local kernels yields only a $2\times$ end-to-end gain because the all-to-all exchange dominates the runtime~\cite{Ayala2020heFFTe}, and follow-up work compresses the exchange lossily to reduce it~\cite{Cayrols2022LossyAlltoall}. Adding GPUs divides the dense working set across devices but does not shrink it, and neither the storage nor the communication decreases with input sparsity. These systems fit workloads that require every coefficient. Spectral signatures do not, and Section~\ref{sec:method} presents methods that exploit sparse input and bound the output instead.

\section{Our Methods}
\label{sec:method}

\begingroup\emergencystretch=2em
We explore three methods that scale the FFT to large sparse matrices. The Binary-Sparse FFT (BS-FFT) computes the exact dense spectrum from sparse input through a custom compressed sparse column (CSC)-based FFT kernel and serves as the lossless reference. Its compute and input storage scale with the nonzero count, so the full dense output remains the dominant cost. The two other methods reduce this output cost in different domains. Elastic BS-FFT evaluates only a requested Cartesian subset of coefficients, so the output size scales with the requested grid. Density-map compression aggregates the input into a coarser grid before the FFT, which reduces input storage, FFT computation, and output storage costs together but does not preserve exact coefficients.

\subsection{BS-FFT: Lossless FFT on Sparse Binary Matrices}
\label{sec:exact-cuspfft}

BS-FFT computes the exact 2-D DFT of a binary sparse matrix $A\in\{0,1\}^{m\times n}$ with $K$ nonzeros, working only from its nonzero entries. Because $A$ is sparse, only those $K$ positions contribute, and BS-FFT evaluates the transform from them without forming the dense $m\times n$ grid. Because $A$ is binary, the transform uses no input values, so each coefficient is a sum of unit-magnitude terms and needs no per-term multiply. The output still stays dense. Sparse input does not give a sparse spectrum, because every nonzero adds one term to every coefficient.

The rows $i$ and columns $j$ index the input, and $u$ and $v$ index the output frequencies. Since $A$ is binary, the DFT of Eq.~\ref{eq:dft2} sums a unit-magnitude term over the nonzero entries
\vspace{-0.3\baselineskip}
\begin{equation}
F[u,v]=\sum_{i,j\,:\,A[i,j]=1} e^{-2\pi\mathbf{i}(ui/m+vj/n)}.
\label{eq:nnz-sum}
\end{equation}
BS-FFT stores these nonzero coordinates as a compressed sparse column (CSC) list of indices~\cite{saad2003iterative}, with no value array, packed to 16-bit integers when $m,n\le 2^{16}$. It keeps only columns with nonzeros, following doubly compressed sparse storage~\cite{buluc2008hypersparse}.

BS-FFT evaluates Eq.~\ref{eq:nnz-sum} in two stages by \textcolor{black}{grouping the terms} over each axis,
\begin{equation}
F[u,v]=\sum_{j}\Big(\sum_{i\,:\,A[i,j]=1} e^{-2\pi\mathbf{i}\,ui/m}\Big)e^{-2\pi\mathbf{i}\,vj/n}.
\label{eq:separable}
\end{equation}
The inner sum reads only the nonzeros of column $j$. For each active column and each frequency $u$, one thread block accumulates it, at $O(K)$ total cost and with no dense grid.

Because $A$ is real, its DFT is Hermitian~\cite{oppenheim1999discrete,sorensen1987real}:
\begin{equation}
F[(-u)\bmod m,(-v)\bmod n]=\overline{F[u,v]}.
\label{eq:hermitian-symmetry}
\end{equation}
Conjugation negates each exponent in Eq.~\ref{eq:nnz-sum} and does not affect the real entries of $A$, so $\overline{F[u,v]}$ equals the same sum evaluated at $(-u,-v)$.

\paragraph{Implementation.}
BS-FFT transforms the columns, the outer sum over $j$, with a 1-D FFT of length $L$. It uses a direct Cooley-Tukey transform with $L=n$ when $n$ factors into 2, 3, 5, and 7, and a Bluestein transform otherwise, which pads $L$ to the smallest such number at or above $2n-1$~\cite{Cooley1965,Bluestein1970,NVIDIAcuFFT}. Because $A$ is real, its spectrum is Hermitian (Eq.~\ref{eq:hermitian-symmetry}), so BS-FFT computes only the first $h=\lfloor m/2\rfloor+1$ frequencies $u$ and fills each mirror row $(m-u)\bmod m$ by conjugation, halving both stages. Algorithm~\ref{alg:cuspfft-gpu} shows the resulting pipeline. BS-FFT processes the $h$ computed frequencies in tiles of $T$ rows. For each tile, a build kernel fills the $T\times L$ signal buffer with the inner sums of Eq.~\ref{eq:separable}, a batched cuFFT transforms the $T$ rows, and a finalize kernel writes each row $u$ and its conjugate mirror row into the output. Two CUDA streams with separate signal buffers alternate tiles, so the FFTs of one tile overlap the build and output writes of the next. Working memory is the two $T\times L$ buffers plus the $m\times q$ output with $q=\lfloor n/2\rfloor+1$, all resident in device memory.

\begin{algorithm}[t]
\caption{BS-FFT lossless transform on the GPU}
\label{alg:cuspfft-gpu}
\begin{algorithmic}[1]
\Require COO coordinates $\{(x_t,y_t)\}_{t=1}^{K}$, dimensions $m,n$, tile height $T$
\Function{TileTransform}{$\mathrm{csc}(A)$, $P$} \Comment{column transform of one tile of row frequencies $P$}
  \State zero the stream's buffer $S\in\mathbb{C}^{|P|\times L}$
  \ForAll{active columns $j$ and rows $u\in P$ in parallel}
    \State $S[u,j]\gets\sum_{i:A[i,j]=1}e^{-2\pi\mathbf{i}\,ui/m}$ \Comment{chirp-scaled on Bluestein}
  \EndFor
  \State batched length-$L$ cuFFT over the rows of $S$ \Comment{forward, multiply, inverse on Bluestein}
  \State \Return $S$
\EndFunction
\State Build binary active-column CSC from the COO coordinates
\State $h\gets\lfloor m/2\rfloor+1$, $q\gets\lfloor n/2\rfloor+1$
\State Set $L\gets n$ for direct cuFFT; otherwise choose a smooth Bluestein length $L\geq2n-1$ and precompute chirps
\State Prepare cuFFT plans, two streams, and two tile buffers
\For{each index tile $P_\tau\subset\{0,\ldots,h-1\}$, alternating two streams}
  \State $S\gets$ \Call{TileTransform}{$\mathrm{csc}(A)$, $P_\tau$}
  \ForAll{$u\in P_\tau$ and $v\in\{0,\ldots,q-1\}$ in parallel}
    \State $F[u,v]\gets S[u,v]$;\quad $F[(m-u)\bmod m,\,v]\gets\overline{S[u,(n-v)\bmod n]}$
  \EndFor
\EndFor
\State \Return $F$
\end{algorithmic}
\end{algorithm}

BS-FFT removes the dense input and its workspace, but it still returns all $mq$ coefficients. It extends the range of feasible exact 2-D FFTs, yet reaches a memory limit on the largest matrices. The next two methods avoid the full output. Elastic BS-FFT (Section~\ref{sec:spectrum-sampling}) calls the same \textsc{TileTransform} as Algorithm~\ref{alg:cuspfft-gpu} and therefore reuses the build kernel, the column FFT, and the tile pipeline.

\subsection{Elastic BS-FFT: Spectrum Sampling}
\label{sec:spectrum-sampling}
\textcolor{black}{Although} BS-FFT utilizes sparse input, it returns the dense spectrum, so its output bounds the reachable matrix size. Elastic Binary-Sparse FFT (Elastic BS-FFT) keeps the BS-FFT pipeline and the original structure of $A$ but computes only requested frequency coefficients. For an $m_0\times n_0$ output, the method returns
\begin{equation}
\widehat F_{\Omega}[p,q]
=\sum_{(i,j)\in\mathrm{nnz}(A)}
e^{-2\pi\mathbf{i}(\bar u_p i/m+\bar v_q j/n)},
\label{eq:sampled-coefficient}
\end{equation}
where $\Omega=U\times V$, $|U|=m_0$, $|V|=n_0$, and $\bar u_p,\bar v_q$ are DFT indices. We select approximately uniform FFT-shifted indices $U$ and $V$ independently along the two axes and evaluate their Cartesian product.

For a block size $b$ we define the requested per-axis sampling rate $r=1/b$, shared with density compression in Section~\ref{sec:density-compression}, and set $m_0=\lceil m/b\rceil$ and $n_0=\lceil n/b\rceil$. Elastic BS-FFT therefore computes exactly $m_0n_0$ coefficients, matching the number returned by the density-map FFT. Both methods have the same realized area sampling rate $r_{\mathrm{area}}$, although they compress different domains.

\begingroup
\emergencystretch=2em
\paragraph{Uniform versus heuristic sampling.}
Sparse FFT methods commonly exploit a sparse or compressible frequency-domain model to identify significant coefficients through downsampling, hashing, or projection measurements~\cite{Hsieh2014SFFTDT,Wang2017FPSSFT}. Our goal differs: we estimate a complete spectrum where its energy may be diffuse, so we compare energy-directed selection with an evenly distributed Cartesian grid.

For a binary sparse matrix $A\in\mathbb{R}^{m\times n}$, let $P=mn$, $K=\mathrm{nnz}(A)$, and let $d=K/P$ denote its spatial density. Let $\widehat A$ be its unitary 2-D DFT. Parseval's identity~\cite{oppenheim1999discrete} states that $\sum_{\omega}|\widehat A_{\omega}|^2=\|A\|_F^2$; hence $q_{\omega}=|\widehat A_{\omega}|^2/\|A\|_F^2$ is a probability distribution over the $P$ frequencies. We use normalized spectral entropy
\begin{equation}
\overline H(q)=
\frac{-\sum_{\omega}q_{\omega}\log q_{\omega}}{\log P}
\in[0,1].
\label{eq:normalized-power-entropy}
\end{equation}
Because $|\widehat A_{\omega}|\leq\|A\|_1/\sqrt{P}
=\sqrt{K/P}\|A\|_F$, we have $\max_{\omega}q_{\omega}\leq d$. Moreover, $H(q)\geq-\log\max_{\omega}q_{\omega}$~\cite{cover1991elements}, which yields
\begin{equation}
\overline H(q)\geq
\frac{\log(1/d)}{\log P}
=1-\frac{\log K}{\log P}.
\label{eq:sparsity-entropy-bound}
\end{equation}
The bound depends on how $K$ grows with $P$. More generally, if $K=O(P^\alpha)$ for $0\leq\alpha<1$, then $\liminf_{P\rightarrow\infty}\overline H(q)\geq1-\alpha$; a lower growth exponent, equivalently a faster decrease in density, raises the entropy floor. Although structured sparse patterns can still contain low-entropy spectral lines, the 15 matrices in Table~\ref{tab:gnn-datasets} exhibit high spectral entropy: the median is $0.978$, with a range of $0.948$ to $0.982$. 

When spectral energy is diffuse, selecting the largest coefficients may retain little more information than uniform sampling at the same sampling rate. Let $p$ be the normalized non-DC power over $M=P-1$ frequencies. The spectral entropy is $H(p)=-\sum_{\omega\neq0}p_\omega\log p_\omega$. It reaches its maximum $\log M$ for the uniform distribution $u_\omega=1/M$~\cite{cover1991elements}. With normalized entropy $\overline H(p)=H(p)/\log M$, define the entropy deficit
\begin{equation*}
\begin{aligned}
\delta&=D_{\mathrm{KL}}(p\|u)=\log M-H(p)\\
&=(1-\overline H(p))\log M.
\end{aligned}
\end{equation*}

For a set $\Omega$ of $Q$ frequencies, write $R(\Omega)=\sum_{\omega\in\Omega}p_\omega$. Pinsker's inequality~\cite{Canonne2022KLTV} gives $|R(\Omega)-Q/M|\leq\sqrt{\delta/2}$. Thus, if $\Omega_H$ contains the $Q$ largest-magnitude coefficients and $\Omega_U$ is a uniform set of the same size
\vspace{-0.3\baselineskip}
\begin{equation}
0\leq R(\Omega_H)-R(\Omega_U)
\leq\sqrt{2\delta}.
\label{eq:retained-energy-gap}
\end{equation}
\textcolor{black}{Since an entropy value close to one indicates} $\delta\to0$, both methods retain approximately $Q/M$ of the energy. This comparison uses an oracle high-magnitude set, so it bounds the possible energy advantage of a heuristic selector. Practical heuristic-based coefficient selection can introduce further error. For example, FPS-SFT~\cite{Wang2017FPSSFT} recovers coordinates and amplitudes from projection bins that pass a one-sparse test, using phase differences between shifted lines. When several frequencies contribute to the same bin, a significant coefficient can be missed. Thus, the selected set may not contain the true largest-magnitude coefficients.

\textcolor{black}{In such a case}, the coverage of the frequency plane could be more important. A heuristic that concentrates samples near high-magnitude coefficients can leave other regions sparsely observed or entirely unsampled, missing local peaks, ridges, or directional structure there. On the normalized frequency plane $\mathcal D$, this coverage can be measured by the largest distance to a sampled coordinate
\vspace{-0.3\baselineskip}
\begin{equation}
h(\Omega)=\sup_{x\in\mathcal D}
\min_{\omega\in\Omega}\|x-\omega\|_2.
\label{eq:sampling-coverage}
\end{equation}
Our uniform sampling has $h(\Omega_U)=O(Q^{-1/2})$, whereas heuristic sampling can leave substantially larger gaps. Uniform sampling therefore provides broader coverage and neighboring values for the bilinear interpolation~\cite{Getreuer2011LinearInterpolation}.

Experiments in Section~\ref{sec:reconstruction-quality} confirm the advantage of our uniform sampling on the high-entropy large sparse matrices.
\par\endgroup

\paragraph{GPU implementation.}
The CUDA implementation follows the BS-FFT pipeline in Section~\ref{sec:exact-cuspfft}. As shown \textcolor{black}{\hypersetup{allcolors=black}in Algorithm~\ref{alg:spectrum-sampling-gpu}}, it builds the compact active-column CSC format (line 1) and reuses \textsc{TileTransform} from Algorithm~\ref{alg:cuspfft-gpu} for the sparse one-dimensional aggregation and the batched column transforms of each tile of sampled frequencies (line 4). Finally, it gathers the requested frequency columns from each row (lines 5--6).

We process the selected-$U$ rows in tiles using two CUDA streams and two memory buffers. While one stream performs the batched cuFFT or Bluestein transforms and gathers the requested $V$ bins, the other stream can process the next tile when GPU resources permit. Each buffer is recycled after its tile completes, and only the gathered $k_u\times k_v$ coefficients are written to persistent output. This approach requires only two $T\times L$ tiles stored in GPU memory, rather than keeping the complete $m\times n$ spectrum.

\begin{algorithm}[t]
\caption{Elastic BS-FFT spectrum sampling on the GPU}
\label{alg:spectrum-sampling-gpu}
\begin{algorithmic}[1]
\Require COO coordinates $\{(x_t,y_t)\}_{t=1}^{K}$, dimensions $m,n$, sampled frequency axes $U=\{\bar u_p\}_{p=1}^{m_0}$ and $V=\{\bar v_q\}_{q=1}^{n_0}$ defining $\Omega=U\times V$, tile height $T$
\State Build binary active-column CSC from the COO coordinates
\State Set $L$ and chirps as in Algorithm~\ref{alg:cuspfft-gpu}; prepare cuFFT plans, two streams, and two tile buffers
\For{each index tile $P_\tau\subset\{\bar u_1,\ldots,\bar u_{m_0}\}$, alternating two streams}
  \State $S\gets$ \Call{TileTransform}{$\mathrm{csc}(A)$, $P_\tau$} \Comment{Algorithm~\ref{alg:cuspfft-gpu}}
  \ForAll{$p\in P_\tau$ and $q\leq n_0$ in parallel}
    \State $\widehat F_\Omega[p,q]\gets S[p,\bar v_q]$ \Comment{gather requested columns}
  \EndFor
\EndFor
\State \Return $\widehat F_\Omega$
\end{algorithmic}
\end{algorithm}

\subsection{Density Map: Spatial-Domain Compression}
\label{sec:density-compression}

Both BS-FFT variants execute the transform pipeline of Algorithm~\ref{alg:cuspfft-gpu}. The density map instead compresses in the spatial domain before any transform, trading exactness for the lowest cost. Prior work~\cite{zhang2026spectralanalysissparsematrix} shows that a density map can compress a sparse pattern into a lower-resolution spatial signal. Let $b\geq1$ be the integer spatial block size and $r=1/b$ the requested per-axis sampling rate. We partition $A\in\{0,1\}^{m\times n}$ into $m_0=\lceil m/b\rceil$ by $n_0=\lceil n/b\rceil$ blocks. Let $B_{pq}$ be block $(p,q)$, $a_{pq}=|B_{pq}|$ its element count, and $C[p,q]=\sum_{(i,j)\in B_{pq}}A[i,j]$ its nonzero count. We form the density map $D\in\mathbb{R}^{m_0\times n_0}$ as
\begin{equation}
\begin{aligned}
D_0[p,q]&=\frac{C[p,q]}{a_{pq}}, &
D[p,q]&=\gamma D_0[p,q],\\
\gamma&=\frac{K}{\sum_{p,q}D_0[p,q]},
\end{aligned}
\label{eq:density-map}
\end{equation}
where $K=\mathrm{nnz}(A)$. The global factor $\gamma$ makes $\sum_{p,q}D[p,q]=K$, so the DC magnitude matches that of the binary input.

Let $\mathcal{F}_{m_0,n_0}$ denote the $m_0\times n_0$ DFT. The compact density-spectrum estimate is
\begin{equation}
M_D^{\Omega}=\left|\operatorname{fftshift}
\!\left(\mathcal{F}_{m_0,n_0}D\right)\right|.
\label{eq:density-estimator}
\end{equation}

\paragraph{Compression ratio versus accuracy.}
Because Density Map compresses the spatial input before the FFT, it introduces spatial aggregation error in addition to any subsequent spectrum-reconstruction error. To isolate the information discarded before the FFT, let $\Pi_bA$ be the blockwise-constant projection that replaces every entry in block $B_{pq}$ by its mean $\pi_{pq}=C[p,q]/a_{pq}$, and define the aggregation loss $E_b^2=\|A-\Pi_bA\|_F^2$. For a binary sparse input $A$,
\begin{equation}
E_b^2=\sum_{p,q}a_{pq}\pi_{pq}(1-\pi_{pq}).
\label{eq:density-projection-error}
\end{equation}
Let $P=mn$, $0<K=\mathrm{nnz}(A)<P$, $d=K/P$, and $Q=m_0n_0$, with area sampling rate $\eta=Q/P=r_{\mathrm{area}}$. Suppose the $K$ occupied positions are uniformly random. Given the block counts $C$, every position within a block has the same conditional mean, so $\mathbb{E}[A\mid C]=\Pi_bA$. The conditional mean minimizes expected squared reconstruction error~\cite{bertsekas2002introduction}. Using the hypergeometric variance of each block count gives
\vspace{-0.3\baselineskip}
\begin{equation}
\frac{\mathbb{E}[E_b^2]}{\|A\|_F^2}
=(1-d)\frac{P-Q}{P-1}
=(1-d)\frac{P}{P-1}(1-\eta).
\label{eq:density-random-rate}
\end{equation}
This is the normalized mean-squared error caused by block averaging; it decreases as the area sampling rate increases.

The same loss has a direct frequency-domain interpretation. Let $F=\mathcal{F}_{m,n}A$ be the unnormalized DFT and let $F_{\neg0}$ omit DC. Since $F[0,0]=K$, Parseval's identity~\cite{oppenheim1999discrete} gives $\|F_{\neg0}\|_2^2=K(P-K)$. The best complex-spectrum estimate based only on $C$ is $\mathcal{F}_{m,n}(\Pi_bA)$, and its normalized mean-squared error is
\vspace{-0.4\baselineskip}
\begin{equation}
\begin{split}
\mathcal{E}_{\mathrm{opt}}(b)
&=\inf_{\widehat F=\widehat F(C)}
\frac{\mathbb{E}\|\widehat F_{\neg0}-F_{\neg0}\|_2^2}
     {K(P-K)}\\
&=\frac{P-Q}{P-1}
=\frac{P}{P-1}(1-\eta).
\end{split}
\label{eq:density-optimal-spectrum-error}
\end{equation}
For divisible dimensions, the sampling rate is $\eta=1/b^2$, so
\begin{equation}
\mathcal{E}_{\mathrm{opt}}(b)
=\frac{P}{P-1}\left(1-\frac{1}{b^2}\right)
\approx 1-\eta.
\label{eq:density-block-rate-error}
\end{equation}
Thus, the minimum normalized mean-squared error of the complex spectrum decreases linearly with the sampling rate $\eta$, approximately following $1-\eta$ for large $P$. The error is zero at $b=1$ and reaches one when the entire input is compressed into a single block.

Note that this calculation does not account for interpolation error when reconstructing the full magnitude spectrum from the compact output. The analysis shows that a higher sampling rate reduces the error caused by spatial compression. However, the reconstruction error need not decrease linearly because interpolation introduces additional error. Adjusting the block size provides a direct way to balance cost and accuracy. Given this potentially nonlinear relationship, an appropriate block size should be selected empirically.

\paragraph{GPU implementation.}
Algorithm~\ref{alg:density-map-gpu} summarizes the implementation of density-map compression \textcolor{black}{on a sparse matrix in COO format}. All intermediates remain on the GPU, avoiding a host round trip between compression and cuFFT.

\begin{algorithm}[t]
\caption{Density-map compression on the GPU}
\label{alg:density-map-gpu}
\begin{algorithmic}[1]
\Require COO coordinates $\{(x_t,y_t)\}_{t=1}^{K}$, dimensions $m,n$, block size $b$
\State $m_0\gets\lceil m/b\rceil$, $n_0\gets\lceil n/b\rceil$
\State Allocate $C\in\mathbb{R}^{m_0\times n_0}$ and initialize it to zero
\ForAll{$t\in\{1,\ldots,K\}$ in parallel}
  \State $p\gets\lfloor x_t/b\rfloor$, $q\gets\lfloor y_t/b\rfloor$
  \State $\operatorname{atomicAdd}(C[p,q],1)$
\EndFor
\ForAll{$(p,q)$ in parallel}
  \State $a_{pq}\gets\min\{b,m-pb\}\min\{b,n-qb\}$ \Comment{valid cells in block $(p,q)$}
  \State $D_0[p,q]\gets C[p,q]/a_{pq}$
\EndFor
\State $\gamma\gets K/\sum_{p,q}D_0[p,q]$; \quad $D\gets\gamma D_0$
\State $M_D^\Omega\gets\left|\operatorname{fftshift}(\operatorname{cuFFT2}(D))\right|$
\State \Return $M_D^\Omega$
\end{algorithmic}
\end{algorithm}

\subsection{Complexity and Memory Comparison}
\label{sec:complexity-memory}

\begingroup\emergencystretch=2em
Let $A$ be an $m\times n$ sparse matrix, $K=\mathrm{nnz}(A)$, and $q=\lfloor n/2\rfloor+1$. Spatial sparsity does not imply a sparse frequency-domain output, so both exact methods must ultimately represent $mq$ nonredundant complex coefficients. We use \emph{work} to denote the computational cost, and \emph{explicit storage} to denote the algorithm-owned arrays, excluding library workspace. Equation~\ref{eq:cuspfft-device-memory} gives their byte counts for BS-FFT and adds the cuFFT workspace $W_S$.
\par\endgroup

\paragraph{Dense cuFFT}
Dense cuFFT materializes the $m\times n$ spatial input and applies the 2-D FFT. Its work is $O(mn\log (mn))$, and its memory cost is $O(mn)$, independent of $K$.

\paragraph{BS-FFT}
The transform of Section~\ref{sec:exact-cuspfft} keeps the compact input, two tile buffers, and the half-spectrum output in device memory. \textcolor{black}{Its work is $O(hK+hL\log L+mq)$, where $h=\lfloor m/2\rfloor+1$ rows are computed and the last term counts the writes of the full output. Conjugate symmetry halves the first two terms but not the third.} For explicit storage, let $a$ be the number of active columns, $\sigma=4\lceil q/4\rceil$ the padded output stride, and $p$ the coordinate width, 2 bytes when both dimensions fit in 16 bits and 4 otherwise. With tile height $T$ and FFT length $L$ from Algorithm~\ref{alg:cuspfft-gpu}, the device allocation is
\begin{equation}
\begin{split}
M_{\mathrm{BSFFT}}^{\mathrm{GPU}}
={}&pK+(p+4)a+4\\
&+8\left(n+L+2TL+m\sigma\right)+W_S(L,T),
\end{split}
\label{eq:cuspfft-device-memory}
\end{equation}
where $2TL$ covers the two tile buffers, $m\sigma$ the output, and $W_S$ the peak cuFFT workspace. The output term $8m\sigma\approx4mn$ dominates, which is about half of the dense arrays because BS-FFT does not form the spatial grid. Equation~\ref{eq:cuspfft-device-memory} predicts the \textcolor{black}{40.068~GB} peak measured on the $100{,}000^2$ \textsf{igb\_tiny}~\cite{Khatua2023IGB}. Dense cuFFT also allocates a 2-D plan workspace, so the memory savings measured in Section~\ref{sec:lossless-perf} are larger.

\paragraph{Elastic BS-FFT}
\begingroup\emergencystretch=2em
With tile height $T$ and FFT length $L$, Elastic BS-FFT requires $O(m_0K+m_0L\log L+m_0n_0)$ work and $O(K+L+2TL+m_0n_0)$ explicit storage. Compared with BS-FFT, it reduces the sparse aggregation and row FFTs to $m_0$ selected rows and the stored output to $m_0n_0$ coefficients, while retaining the sparse input and tile buffers.
\par\endgroup

\paragraph{\textcolor{black}{Density Map}.}
For an $m_0\times n_0$ output, density compression first scans the $K$ nonzeros and then transforms the reduced spatial grid. Its work is $O(K+m_0n_0(\log m_0+\log n_0))$, and its explicit storage is $O(K+m_0n_0)$. It is the least expensive because no operation is performed at the original size.

\subsection{Approximate Full-Spectrum Magnitude Reconstruction}
\label{sec:full-spectrum-reconstruction}
\begingroup
\emergencystretch=2em

Density Map and \textcolor{black}{Elastic BS-FFT} return $m_0n_0$ values rather than the full $m\times n$ spectrum. For visualization and reconstruction-quality assessment, we recover a full-resolution magnitude image by bilinear interpolation~\cite{Getreuer2011LinearInterpolation}. After the reduced FFT, Density Map maps its compact magnitude grid onto the full-resolution shifted frequency image by aligning the endpoints along each axis. For $m_0,n_0>1$, the mapped node positions are:
\vspace{-0.3\baselineskip}
\begin{equation}
x_p^D=\frac{p(m-1)}{m_0-1},
\qquad
y_q^D=\frac{q(n-1)}{n_0-1}.
\label{eq:density-resize-coordinates}
\end{equation}
\textcolor{black}{Elastic BS-FFT} instead uses the actual shifted DFT coordinates $x_p^S,y_q^S$ of its sampled coefficients. With these node positions defined, both methods use the same bilinear interpolation~\cite{Getreuer2011LinearInterpolation}
described below, \textcolor{black}{since it fits their rectangular grids and keeps interpolated
magnitudes within the range of neighboring samples.}

Let $Z_{pq}$ denote either $M_D^\Omega[p,q]$ at $(x_p^D,y_q^D)$ or $|\widehat F_\Omega[p,q]|$ at $(x_p^S,y_q^S)$. For a target coordinate $(x,y)$ bracketed by rows $p,p+1$ and columns $q,q+1$, define
\vspace{-0.4\baselineskip}
\begin{equation}
\alpha=\frac{x-x_p}{x_{p+1}-x_p},
\qquad
\beta=\frac{y-y_q}{y_{q+1}-y_q}.
\end{equation}
The reconstructed magnitude is
\begin{equation}
\begin{split}
\widetilde M[x,y]={}&(1-\alpha)(1-\beta)Z_{p,q}
 +(1-\alpha)\beta Z_{p,q+1}\\
&+\alpha(1-\beta)Z_{p+1,q}
 +\alpha\beta Z_{p+1,q+1},
\end{split}
\label{eq:full-magnitude-reconstruction}
\end{equation}

The DC value is calculated separately
\begin{equation}
F_A[0,0]=\sum_{i=0}^{m-1}\sum_{j=0}^{n-1}A[i,j]=K.
\label{eq:known-dc}
\end{equation}
If a complex reconstruction is required, the same weights can be applied separately to sampled real and imaginary parts.

Both decoders materialize $mn$ values and therefore require $\Theta(mn)$ work and storage. Note that full-spectrum reconstruction isn't always necessary: when a user needs only spectral features, decoders in Section~\ref{sec:direct-features} operates directly on the compact output.
\par\endgroup

\subsection{Spectral Signature Decoder}
\label{sec:direct-features}

\begingroup\emergencystretch=2em
Expanding a compact spectrum to all $mn$ frequencies would reintroduce $\Theta(mn)$ interpolation work and storage, defeating the memory-saving design. \textcolor{black}{Many applications use compact spectral descriptors: spectral entropy for speech endpoint detection~\cite{Shen1998SpectralEntropy}, pooled spectral responses for scene recognition~\cite{Oliva2001SpatialEnvelope}, and Gabor-feature statistics for texture retrieval~\cite{Manjunath1996TextureFeatures}.} We therefore propose to compute the final features directly from the compact spectrum.
\par\endgroup

Let $\mathcal{S}=\{(\boldsymbol{\omega}_t,Z_t,\alpha_t)\}_{t=1}^{Q}$, where $Q=m_0n_0$, $Z_t$ is a retained complex coefficient, $\boldsymbol{\omega}_t=(u_t/m,v_t/n)$ is its normalized shifted coordinate, and $\alpha_t$ is the number of original frequency cells represented by the sample. For the evenly distributed Cartesian grid, $\alpha_t=mn/Q$ for every $t$; an irregular native support instead uses the area of its clipped frequency-domain Voronoi cell so that densely sampled regions do not receive disproportionate weight.

Due to space constraints, we use entropy and radial energy to illustrate the spectral signature decoding. Define the represented energy, total energy, and radius as
\vspace{-0.3\baselineskip}
\begin{equation}
\begin{aligned}
e_t&=\alpha_t|Z_t|^2,&
S&=\sum_{t=1}^{Q}e_t,\\
r_t&=\|\boldsymbol{\omega}_t\|_2.
\end{aligned}
\label{eq:direct-feature-energy}
\end{equation}
For $S>0$, the normalized energy mass is
\begin{equation}
p_t=\frac{\alpha_t|Z_t|^2}
          {\sum_{j=1}^{Q}\alpha_j|Z_j|^2}
=\frac{|Z_t|^2}{\sum_{j=1}^{Q}|Z_j|^2}
\quad\text{when }\alpha_t\equiv mn/Q.
\label{eq:direct-feature-normalization}
\end{equation}
We then compute normalized spectral entropy and radial energy distributions:
\vspace{-0.3\baselineskip}
\begin{equation}
\begin{aligned}
\widehat H&=-\frac{\sum_{t:p_t>0}p_t\log p_t}{\log Q},\\
\widehat R_k&=\sum_{t=1}^{Q}p_t\,
 \mathbf{1}[r_t\in\mathcal{B}_k],
\end{aligned}
\label{eq:direct-radial-features}
\end{equation}
where $\{\mathcal{B}_k\}$ partitions $[0,r_{\max}]$ into radial bins. Other spectral signatures use similar approaches.

The extractor requires one pass over the $Q$ coefficients. After gathering a selected tile, each GPU thread forms one $e_t$ and $r_t$. A shared-memory reduction accumulates $S$, entropy, and the radial-bin totals before each block atomically updates the global accumulators. The resulting cost is $O(Q)$ work and $O(B_r)$ auxiliary storage.

\section{Evaluation}
\label{sec:evaluation}

\begingroup
\emergencystretch=2em
Prior work \textcolor{black}{has already demonstrated the significant benefits of spectrum analysis for various sparse-matrix applications}~\cite{zhang2026spectralanalysissparsematrix}. Therefore, this work focuses on the scalability of the proposed sparse FFT methods. All tasks are run on one NVIDIA A100 40GB GPU. The reported runtime and peak GPU memory usage is the median of 10 runs following one warm-up run. We organize the evaluation around four research questions:
\begin{itemize}
  \item \textbf{RQ1: Scalability and performance.} How well does BS-FFT scale spectral analysis of sparse matrices without information loss?
  \item \textbf{RQ2: Trade-offs and elasticity.} How does varying the sampling rate in Elastic BS-FFT and Density Map balance spectral accuracy against execution time and memory use?
  \item \textbf{RQ3: Overhead analysis.} What are the postprocessing costs of full-spectrum reconstruction and spectral feature decoding?
  \item \textbf{RQ4: Method selection.} How should users choose among the methods in different application scenarios?
\end{itemize}

\subsection{Experimental Setup}

\paragraph{Dataset.}
We use 15 adjacency matrices drawn from established GNN benchmarks and prior GPU sparsity studies~\cite{Mernyei2020WikiCS,Shchur2018Pitfalls,Hu2020OGB,Khatua2023IGB,Chen2025NMSparsity}. Table~\ref{tab:gnn-datasets} lists their dimensions and sparsity. We binarize every matrix because the study concerns sparsity patterns rather than edge weights. 

\begin{table}[t]
  \centering
  \footnotesize
  \setlength{\tabcolsep}{2.4pt}
  \caption{GNN adjacency-matrix dataset.}
  \label{tab:gnn-datasets}
  \begin{tabular}{@{}lrrr@{}}
    \toprule
    Dataset & Nodes & Edges & Density \\
    \midrule
    Wiki-CS~\cite{Mernyei2020WikiCS} & 11,701 & 290,519 & $ 2.12\times10^{-3} $ \\
    Amazon Computers~\cite{Shchur2018Pitfalls} & 13,752 & 491,722 & $ 2.60\times10^{-3} $ \\
    Coauthor CS~\cite{Shchur2018Pitfalls} & 18,333 & 163,788 & $ 4.87\times10^{-4} $ \\
    CoraFull~\cite{Bojchevski2018Graph2Gauss} & 19,793 & 126,842 & $ 3.24\times10^{-4} $ \\
    Facebook Page-Page~\cite{Rozemberczki2019MUSAE} & 22,470 & 341,646 & $ 6.77\times10^{-4} $ \\
    Deezer Europe~\cite{Rozemberczki2020FEATHER} & 28,281 & 185,504 & $ 2.32\times10^{-4} $ \\
    Coauthor Physics~\cite{Shchur2018Pitfalls} & 34,493 & 495,924 & $ 4.17\times10^{-4} $ \\
    GitHub~\cite{Rozemberczki2019MUSAE} & 37,700 & 578,006 & $ 4.07\times10^{-4} $ \\
    Penn94~\cite{Lim2021LINKX} & 41,554 & 2,724,458 & $ 1.58\times10^{-3} $ \\
    PPI~\cite{Hamilton2017GraphSAGE} & 56,944 & 1,587,264 & $ 4.90\times10^{-4} $ \\
    Yelp2018~\cite{He2020LightGCN} & 69,716 & 3,122,812 & $ 6.43\times10^{-4} $ \\
    Gowalla~\cite{He2020LightGCN} & 70,839 & 2,054,740 & $ 4.09\times10^{-4} $ \\
    Flickr~\cite{Zeng2020GraphSAINT} & 89,250 & 899,756 & $ 1.13\times10^{-4} $ \\
    \texttt{ogbl-biokg}~\cite{Hu2020OGB} & 93,773 & 3,540,566 & $ 4.03\times10^{-4} $ \\
    IGB-tiny~\cite{Khatua2023IGB} & 100,000 & 447,076 & $ 4.47\times10^{-5} $ \\
    \bottomrule
  \end{tabular}
\end{table}

\paragraph{Metrics.}
\begingroup\emergencystretch=2em
We report GPU execution time and peak device memory for cost analysis. For compression-based methods, we additionally report reconstruction quality and spectral feature estimation accuracy. Our primary reconstruction metric is the Hellinger distance between the reconstructed and \textcolor{black}{reference full-spectrum normalized magnitude distributions}~\cite{Hellinger1909}. Hellinger distance is widely used in statistics to quantify discrepancies between probability distributions~\cite{GibbsSu2002ProbabilityMetrics}, making it suitable for assessing how closely a reconstructed spectral distribution matches the reference. We form each distribution by squaring the magnitudes and normalizing them to unit total power. For candidate and reference distributions $p$ and $q$, respectively, the distance is
\begin{equation}
H(p,q)=\frac{1}{\sqrt{2}}\left\|\sqrt{p}-\sqrt{q}\right\|_2
=\sqrt{1-\sum_i\sqrt{p_iq_i}}.
\label{eq:hellinger}
\end{equation}
We evaluate the following spectral features:
(1) normalized spectral entropy;
(2) a 16-bin radial energy distribution; and
(3) an eight-sector directional energy distribution. (Definitions in ~\cite{zhang2026spectralanalysissparsematrix} Figure~1)
The radial and directional vectors are normalized and evaluated with the Hellinger distance in Eq.~\ref{eq:hellinger}. Spectral entropy is evaluated by its relative error.

\subsection{Baselines}
\label{sec:baselines}
\textcolor{black}{The baseline methods and libraries we used} in the experiment include the following:
\begingroup\emergencystretch=2em
\begin{itemize}
  \item \textbf{Dense full transforms.} Dense cuFFT~\cite{NVIDIAcuFFT} materializes the spatial input and computes a complete 2-D R2C transform.
  \item \textbf{Spatial baseline.} Fixed-point compression retains fixed input coordinates and applies a dense FFT to the reduced grid.
  \item \textbf{Spectrum baselines.} For coordinate-given execution, we use FINUFFT~\cite{Barnett2019FINUFFT}\textcolor{black}{\hypersetup{allcolors=black} through its CUDA backend, cuFINUFFT~\cite{Shih2021cuFINUFFT},} and SparseFFT.jl~\cite{HoSparseFFTjl}. Note that we adapted \textcolor{black}{SparseFFT.jl} for CUDA execution for fair comparisons, and \textcolor{black}{our measurements show that the adapted GPU kernels are faster than the original Julia-based CPU implementation}. They receive the same Cartesian coordinates as \textcolor{black}{Elastic BS-FFT}. FPS-SFT, on the other hand, chooses its own coordinates from projection-slice measurements~\cite{Wang2017FPSSFT}. It is our native coefficient-selecting baseline. We match its output coefficient count and preserve its selected coordinates.
\end{itemize}
\par\endgroup

\begin{figure}[t]
  \centering
  \includegraphics[width=0.9\columnwidth]{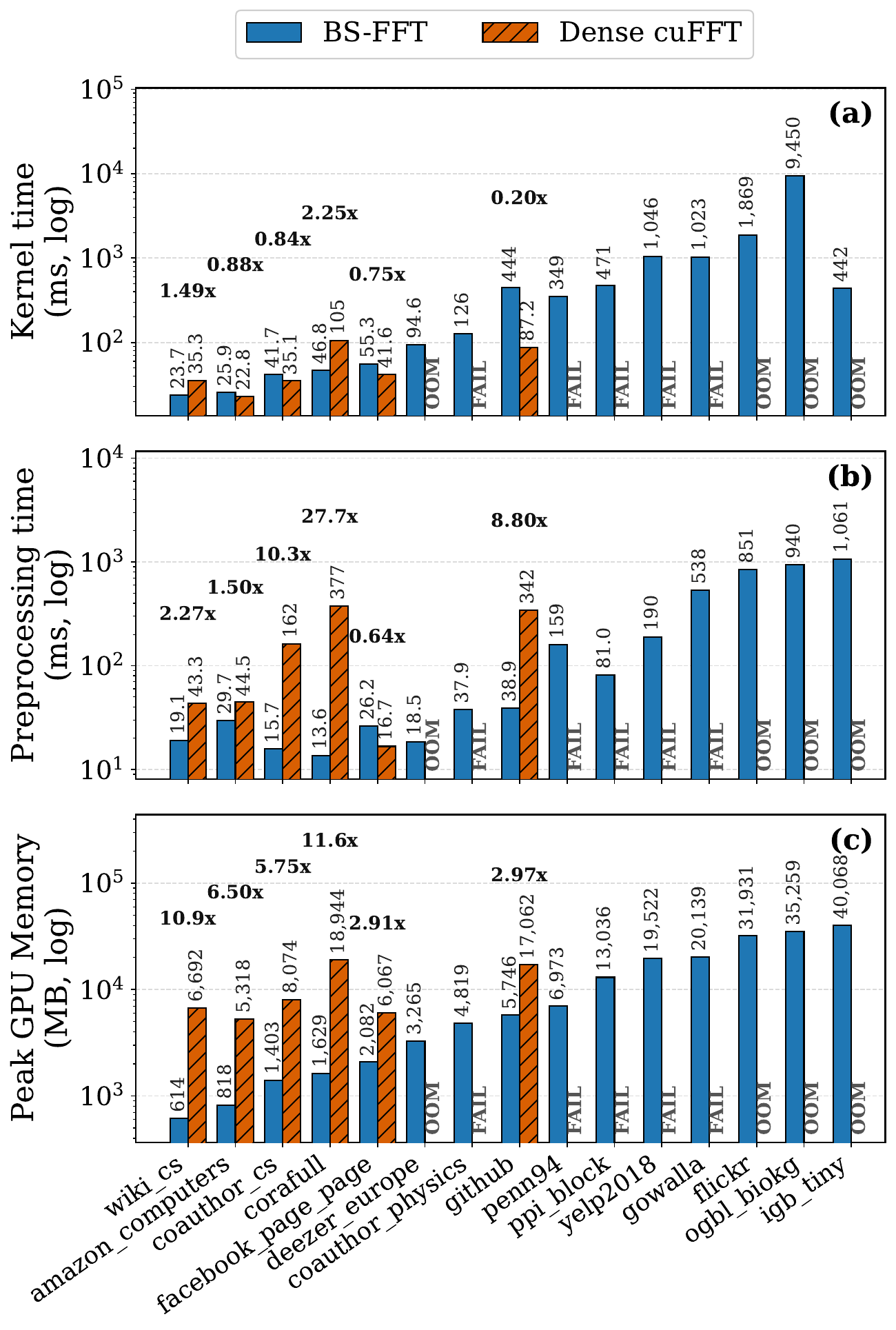}
  \caption{Dense cuFFT and BS-FFT on the 15 GNN adjacency matrices on an A100 40~\textcolor{black}{GB}. (a) GPU kernel time. (b) Host preprocessing time. (c) Peak device memory.}
  \label{fig:gnn15-lossless}
\end{figure}

\subsection{Lossless BS-FFT Performance (RQ1)}
\label{sec:lossless-perf}

We compare dense cuFFT and BS-FFT on the 15 GNN matrices of Table~\ref{tab:gnn-datasets} on one A100 with 40GB of device memory, results are shown in Figure~\ref{fig:gnn15-lossless}. Dense cuFFT completes 6 of the 15 matrices. Four fail at allocation, marked OOM, and five fail at plan creation because a dimension contains a large prime factor, marked FAIL. BS-FFT completes all 15, including the $100{,}000^2$ \textsf{igb\_tiny} at a \textcolor{black}{40.068~GB} peak, and where dense cuFFT runs the two outputs agree to a maximum relative error of $6.7\times10^{-3}$. On the six jointly completed matrices, dense cuFFT has the faster kernel on four because one monolithic 2-D FFT is fast whenever the plan succeeds and the grid fits, while BS-FFT pays the $O(hK)$ row construction of Section~\ref{sec:complexity-memory}. For preprocessing, dense cuFFT scatters the nonzeros into the $mn$ grid and creates a 2-D plan, up to 377~ms on \textsf{corafull}, whereas BS-FFT builds its compact CSC in 14 to 39~ms on the same matrices. Summing both stages, BS-FFT is faster end to end on four of the six, up to $7.98\times$ on \textsf{corafull}, and \textcolor{black}{it reduces peak memory use by 2.9--11.6$\times$} on every jointly completed matrix, consistent with Eq.~\ref{eq:cuspfft-device-memory}.

\subsection{Compression-Based Methods (RQ2)}

\subsubsection{Sampling Rate}
For compression-based methods, we use block side lengths $b\in\{4,8,16,32,64,128\}$ and define the requested per-axis sampling rate as $r=1/b$; the element-wise sampling rate is therefore $1/b^2$. 

\begingroup\emergencystretch=2em
\textcolor{black}{To achieve a fair comparison} among different sampling methods, we use adaptive sampling. For an $m\times n$ matrix, \textcolor{black}{Density Map} and \textcolor{black}{Fixed Point spatial compression methods} interpret $b$ literally: each output cell aggregates one spatial $b\times b$ block, giving $m_0=\lceil m/b\rceil$ and $n_0=\lceil n/b\rceil$. Elastic BS-FFT, FINUFFT, and SparseFFT.jl translate the same $b$ into a matched coefficient budget by selecting $m_0\times n_0$ DFT coordinates on a Cartesian grid. For example, $b=4$ retains approximately $1/16$ of the cells or coefficients. For FPS-SFT, only the output count $m_0n_0$ is matched since it has a separate coordinate-selection rule.
\par\endgroup

\begin{figure}[t]
  \centering
  \includegraphics[width=0.75\columnwidth]{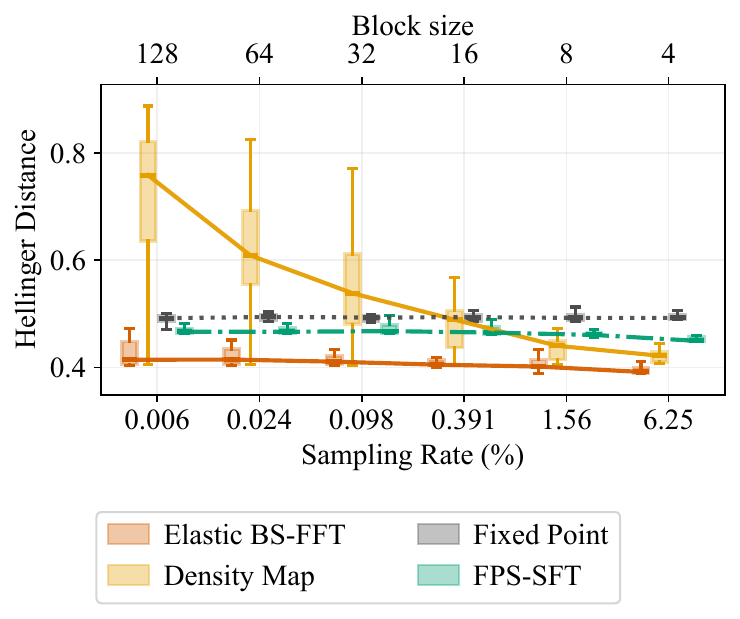}
  \caption{Hellinger distance error of reconstructed versus original spectral magnitude on 15 GNN adjacency matrices.}
  \label{fig:gnn15-full-spectrum-hellinger}
\end{figure}

\begin{figure}[t]
  \centering
  \includegraphics[width=0.97\columnwidth]{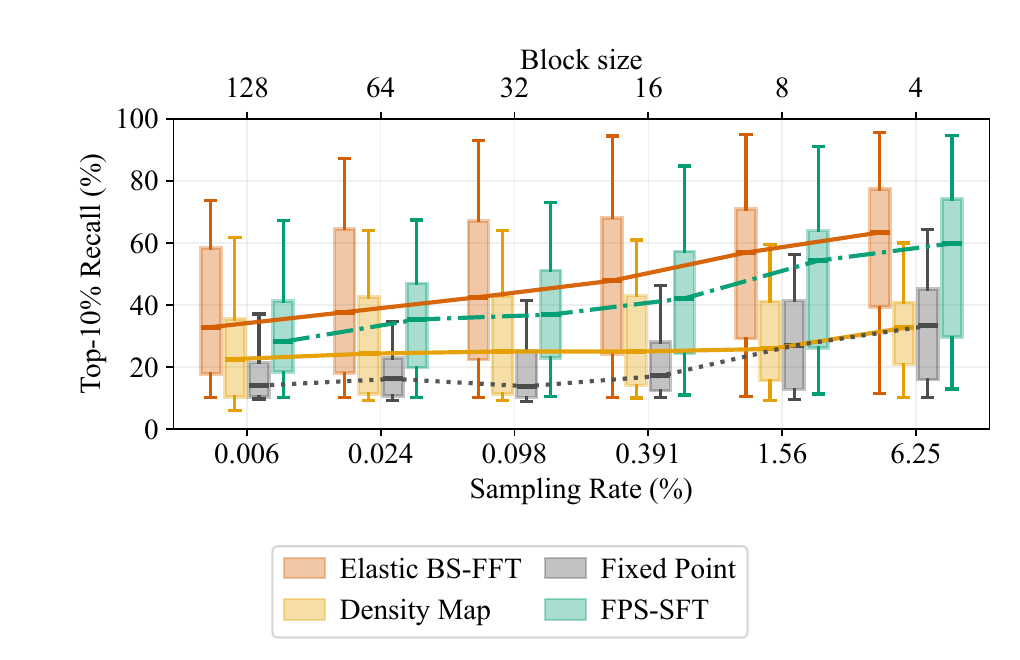}
  \caption{Recovery of dominant spectral components, measured by recall of the top 10\% highest-magnitude coefficients in the reference spectrum on 15 GNN adjacency matrices.}
  \label{fig:ss15-top10-recall}
\end{figure}

\begin{figure}[t]
  \centering
  \includegraphics[width=\columnwidth]{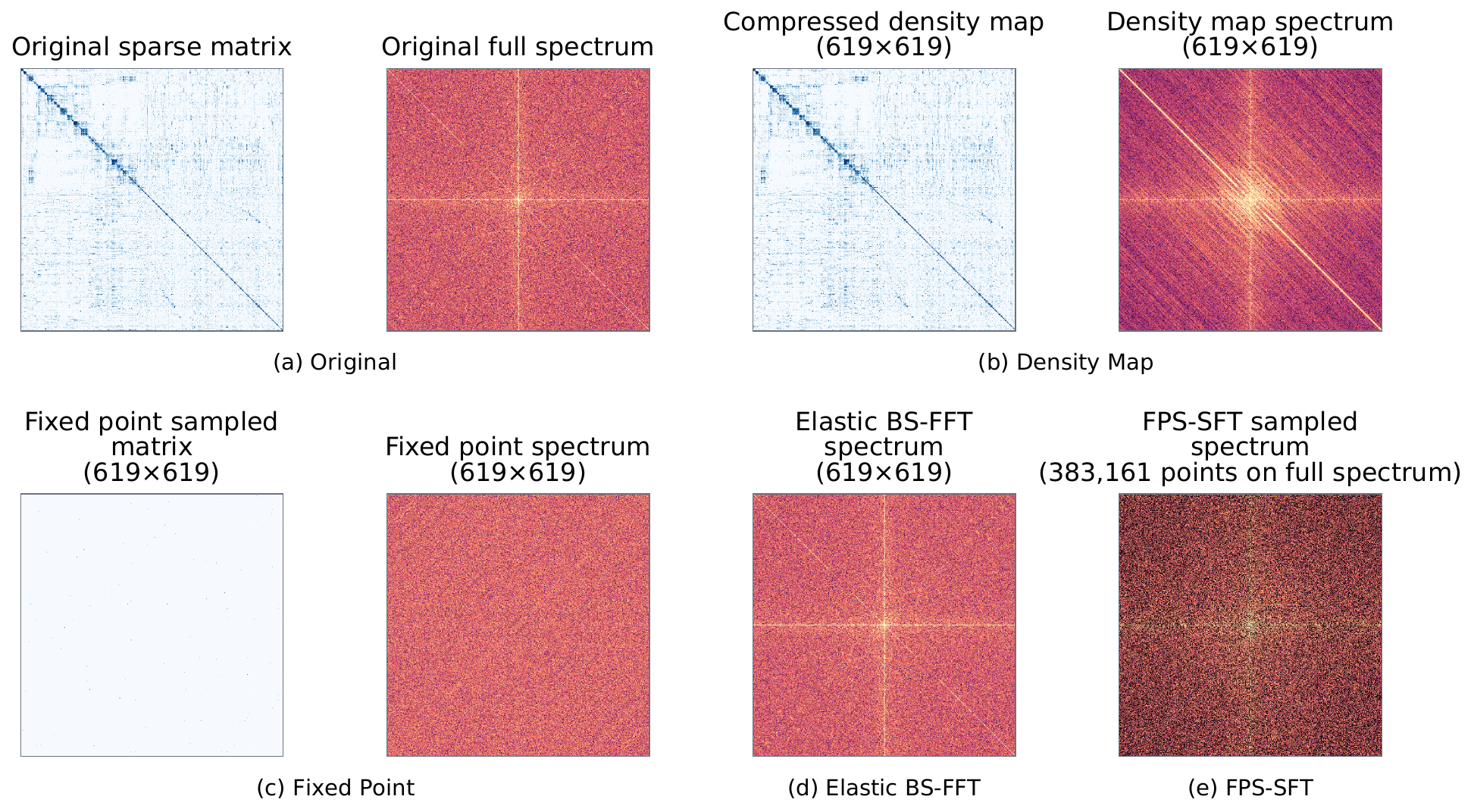}
  \caption{Different compression methods on CoraFull (\(19{,}793\times19{,}793\)) with block size 32 (sampling rate 0.098\%). (a) Original sparse matrix and spectrum. (b) Density Map and its spectrum. (c) Fixed-point and its spectrum. (d) Elastic BS-FFT spectrum on the sampled Cartesian gird. (e) FPS-SFT sampled locations and magnitudes on the full frequency grid.}
  \label{fig:corafull-method-illustration}
\end{figure}

\subsubsection{Compression Quality}
\label{sec:reconstruction-quality}
\begingroup\emergencystretch=2em To show our compression method's effectiveness, we first compare the spectral fidelity of our methods with the baselines for both full-spectrum reconstruction and spectral feature estimation.
\paragraph{Spectrum reconstruction.}
\begingroup\emergencystretch=2em We use methods detailed in Section~\ref{sec:full-spectrum-reconstruction} to \textcolor{black}{reconstruct the full spectrum}. We then compare \textcolor{black}{reconstructed normalized magnitude distributions} with the reference dense spectrum in Figure~\ref{fig:gnn15-full-spectrum-hellinger}. In the box plots, center lines and connecting lines indicate medians, boxes indicate the interquartile range (IQR), and whiskers extend to 1.5~IQR. Elastic BS-FFT has the lowest error at every tested rate, decreasing from 0.414 at 0.006\% sampling to 0.391 at 6.25\%. Density Map is the least accurate \textcolor{black}{at low sampling rates}, dropping from 0.759 to 0.422 as the sampling rate increases. \textcolor{black}{Fixed Point}, on the other hand, remains constant at 0.494, indicating that the full-spectrum estimation quality is not improving with sampling rate. FPS-SFT sits between \textcolor{black}{Fixed Point} and Elastic BS-FFT \textcolor{black}{with error decreasing only slightly from 0.45 to 0.43}. This is likely because the FPS-SFT algorithm is \textcolor{black}{designed for a sparse spectrum}, and if requested to generate more coordinates, it fails to capture more dominant magnitudes and \textcolor{black}{does not} provide much extra information.\par\endgroup

\begingroup\emergencystretch=2em Figure~\ref{fig:corafull-method-illustration} \textcolor{black}{illustrates these differences using CoraFull as a case study}: Density Map retains the structural pattern but \textcolor{black}{concentrates energy at lower frequencies}, and Section~\ref{sec:density-compression} proved that \textcolor{black}{there is} more information loss when \textcolor{black}{the sampling rate} is smaller. \textcolor{black}{Fixed Point} completely loses the \textcolor{black}{structural information} in the spatial domain, and the spectrum is \textcolor{black}{diffuse}, explaining its constant error ratio. Elastic BS-FFT's compressed spectrum looks most similar to the original spectrum. FPS-SFT appears darker because its samples are scattered across the full frequency plane rather than arranged on a compact Cartesian grid. It preserves the \textcolor{black}{cross-shaped high-magnitude structure} in the original spectrum, but fails to highlight the diagonal energy band extending from the upper-left to the lower-right corner.\par\endgroup

\begingroup\emergencystretch=2em \textcolor{black}{Although} the Density Map's compressed spectrum preserves the structural information, the error is high because it causes some frequency shifts. In practice, however, applications may focus on spectrum peaks and patterns, \textcolor{black}{as signal compression focuses on} high-magnitude coefficients~\cite{Hassanieh2012PracticalSFFT}, while scene recognition uses spectral patterns~\cite{Oliva2001SpatialEnvelope}. Full-spectrum Hellinger error does not directly measure the preservation of these locations or patterns. This motivates a separate comparison of how well reconstruction recovers the locations of high-magnitude components.\par\endgroup

\begingroup
\emergencystretch=2em
Figure~\ref{fig:ss15-top10-recall} further evaluates the reconstruction quality by showing the recall rate of the \textcolor{black}{coordinates with the largest 10\% of magnitudes}: let $T$ and $R$ contain the Top-10\% non-DC coordinates of the reference and reconstructed magnitude spectra, respectively. Recall is $|T\cap R|/|T|$, so it measures recovered high energy locations rather than the full spectrum. At $b=16$, average recall is 48.6\% for Elastic BS-FFT, 43.4\% for FPS-SFT, 29.0\% for Density Map, and 21.0\% for \textcolor{black}{Fixed Point}. Density Map has a clear advantage \textcolor{black}{compared with Fixed Point at low sampling rates}, and Elastic BS-FFT can recover more than 60\% of the energy peaks on average at 6.25\% sampling rate.
\par\endgroup

\begin{figure*}[t]
  \centering
  \includegraphics[width=0.92\textwidth]{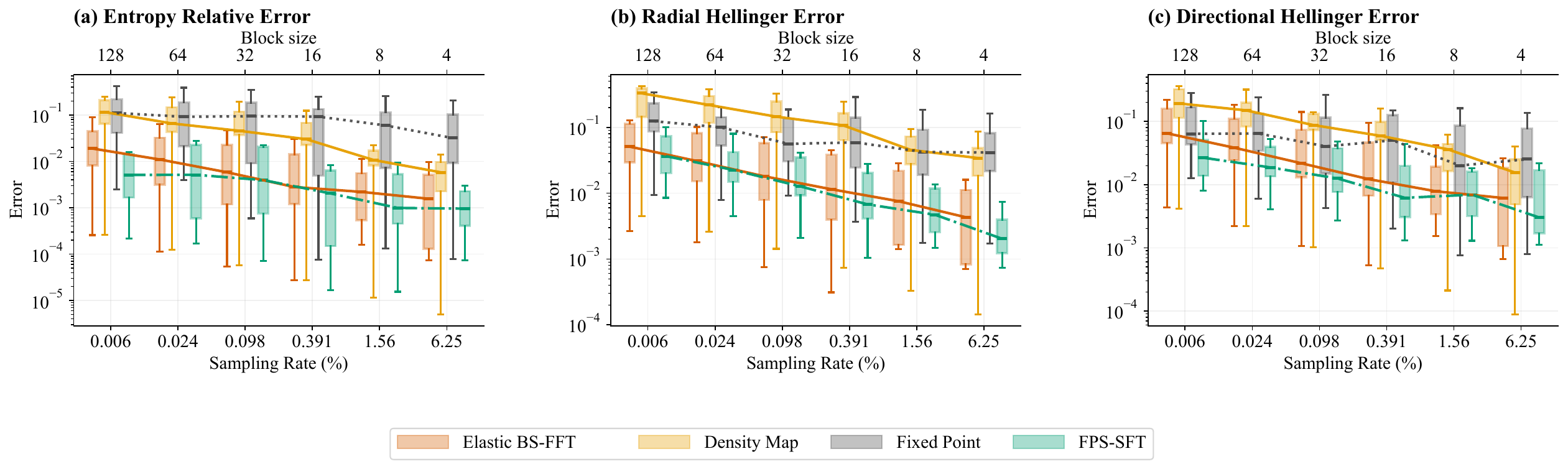}
  \caption{Spectral-feature accuracy across sampling rates on 15 GNN adjacency matrices. (a) Entropy relative error. (b) Radial Hellinger error. (c) Directional Hellinger error.}
  \Description{Three panels compare entropy relative error, radial Hellinger error, and directional Hellinger error for Elastic BS-FFT, Density Map, fixed-point sampling, and FPS-SFT across six sampling rates on GNN-15.}
  \label{fig:gnn15-feature-quality}
\end{figure*}

\begin{figure}[t]
  \centering
  \includegraphics[width=\columnwidth]{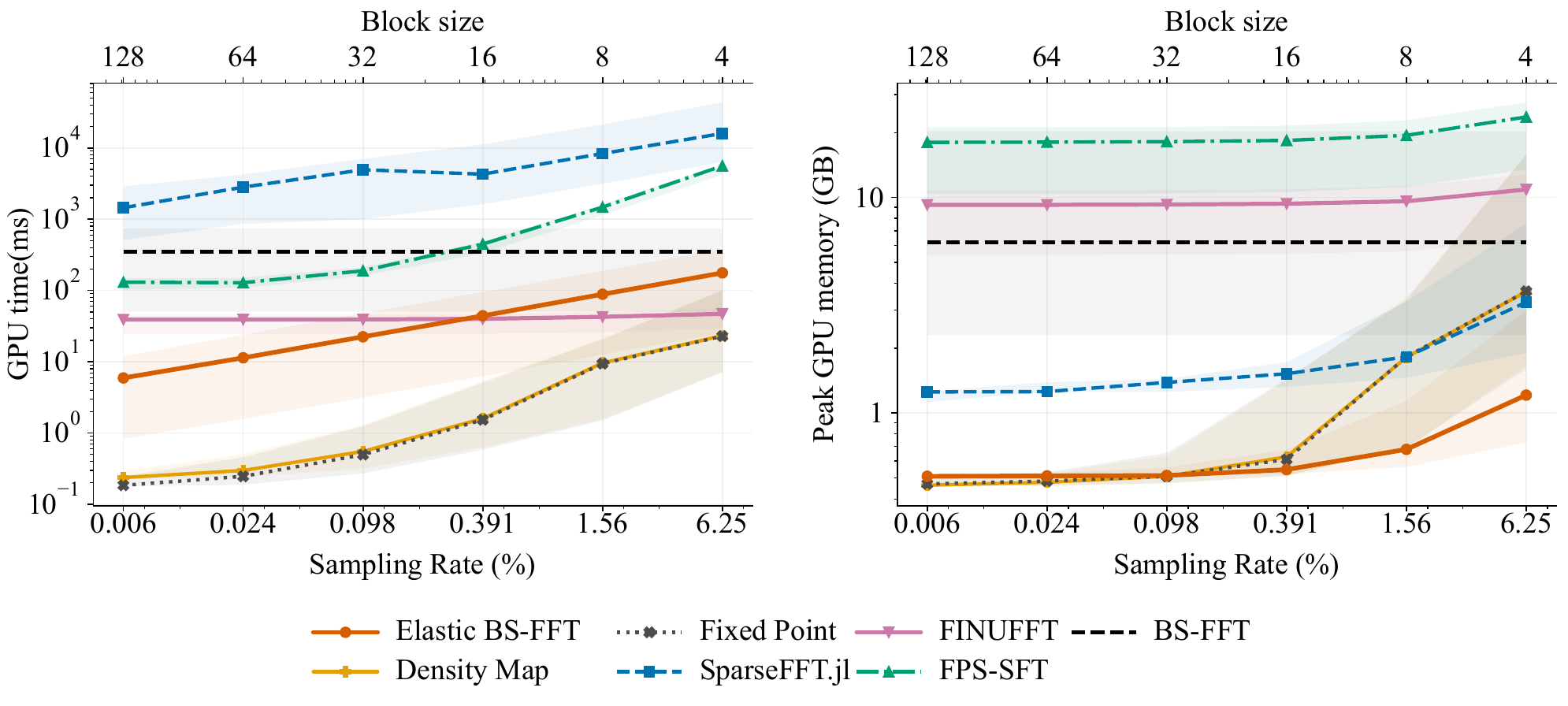}
  \caption{Performance and memory across sampling rates on 15 GNN adjacency matrices. (a) Median A100 GPU time. (b) Median peak GPU memory.}
  \Description{Two panels show median A100 GPU time and peak GPU memory versus sampling rate, with shaded interquartile ranges for seven methods.}
  \label{fig:multirate-pilot-spectral}
\end{figure}

\paragraph{Spectral Signature Estimation.}
\begingroup\emergencystretch=2em
As mentioned in Section~\ref{sec:direct-features}, many applications only use certain spectral features. Figure~\ref{fig:gnn15-feature-quality} \textcolor{black}{evaluates the accuracy of estimating entropy}, radial energy, and directional energy directly from the spectra. Elastic BS-FFT and FPS-SFT \textcolor{black}{have similar accuracy, with median entropy relative error ranging from 0.0016 to 0.0189, radial Hellinger distance from 0.0043 to 0.0509, and directional Hellinger distance from 0.0061 to 0.0647}, indicating they are \textcolor{black}{accurate} in terms of estimating spectral features. \textcolor{black}{Density Map} and Fixed Point, on the other hand, are far less accurate, sometimes exceeding 10\% error in the three spectral features. Moreover, Density Map performs worse on Radial and Directional features because it shifts the energy to the low frequency area\textcolor{black}{; meanwhile,} it has \textcolor{black}{a wide range of error values across the dataset}, showing \textcolor{black}{it is only} suitable for certain matrices with clear sparse patterns. \textcolor{black}{In general, Elastic BS-FFT and FPS-SFT} are more suitable for direct feature estimation, \textcolor{black}{whereas} Density Map and Fixed Point rely on \textcolor{black}{higher sampling rates in order} to be accurate enough.
\par\endgroup

\subsubsection{Computation Cost}
Figure~\ref{fig:multirate-pilot-spectral} compares GPU time and peak device memory of the compression-based methods with dense BS-FFT, whose full-spectrum cost does not vary with sampling rate. Across all matrices, Elastic BS-FFT requires 5.93--176.55 ms and \textcolor{black}{0.51--1.21~GB} of memory, versus 348.299 ms and \textcolor{black}{6.19~GB} for BS-FFT. It is therefore 2.0--58.7$\times$ faster and \textcolor{black}{reduces device memory use by 5.1--12.1$\times$}. Density Map is even faster at 0.238--23.20 \textcolor{black}{ms and consumes less memory} \textcolor{black}{at lower sampling rates,} although its memory grows more steeply. Elastic BS-FFT trades some of Density Map's speed for the stronger spectrum and feature accuracy.

\textcolor{black}{FPS-SFT,} while achieving decent accuracy, is slower than BS-FFT at sampling rates higher than 0.391\% and \textcolor{black}{consistently consumes} more memory. This is because its selected coordinates do not form a GPU-friendly Cartesian grid, and it does not utilize the sparsity in the input matrix. SparseFFT.jl also fails to take advantage of the Cartesian grid output and sparse input, while FINUFFT computes the full subspectrum bounded by the largest frequency coordinates, which is only suitable for cases where energy is centered in the spectrum. They either consume more memory or \textcolor{black}{achieve smaller speedups} compared to BS-FFT, making them ineffective in our task scenario.

\subsection{Postprocessing Cost (RQ3)}
\label{sec:postprocessing-cost}

We also measure postprocessing time on the dataset, including full-spectrum reconstruction and spectral signature decoder. We use the NVIDIA A100 for tiled bilinear interpolation, along with 64 AMD EPYC 7763 CPUs for memory I/O. \textcolor{black}{At sampling rates of 6.25\%, 0.391\%, and 0.006\%}, reconstruction takes an average of 2.886, 2.776, and 2.727~s, with ranges of 0.161--13.3, 0.160--12.9, and 0.161--12.3~s, respectively. The cost changes little with the sampling rate because the full $mn$ output must still be computed and written.

Direct spectral feature extraction yields mean times of 524.08, 42.10, and 4.04~ms, with ranges of 34.36--1652.72, 7.48--115.74, and 1.40--8.16~ms, respectively. This shows that computing only the spectral features is much more efficient and \textcolor{black}{benefits from lower sampling rates}.

\begin{figure}[t]
  \centering
  \includegraphics[width=0.48\textwidth]{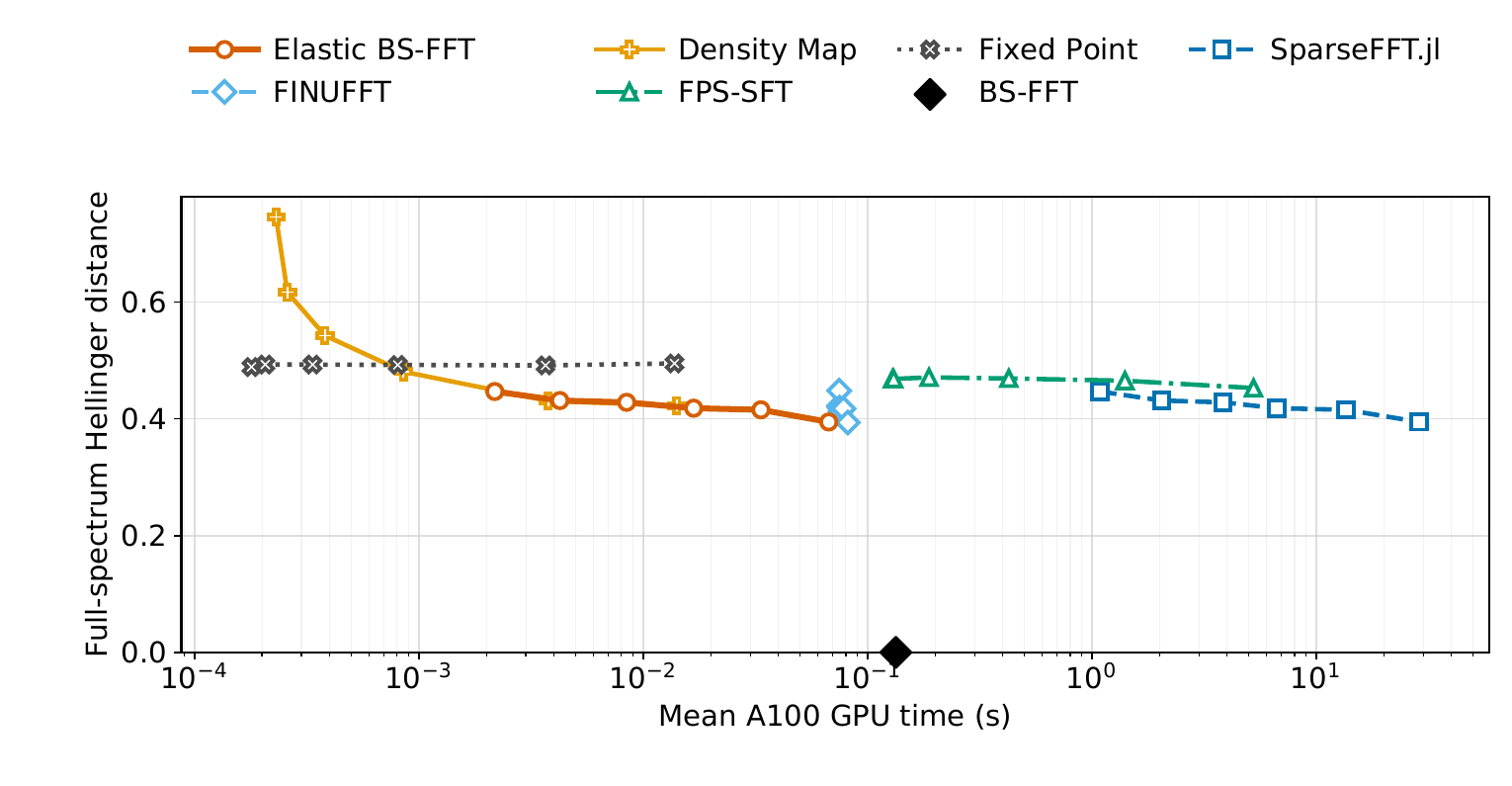}
  \caption{Reconstruction quality, measured by the Hellinger distance between the reconstructed and reference spectra, versus computational cost, measured by GPU runtime. }
  \label{fig:gnn15-hellinger-time-tradeoff}
\end{figure}

\subsection{Choosing a Sparse FFT Method (RQ4)}
\label{sec:method-selection}
The three proposed methods have different strengths and limitations. For relatively small matrices, dense cuFFT remains faster. BS-FFT provides a lossless sparse-input solution, but it still has an output-size limit and can be relatively costly. It is most useful when the matrix fits within its memory bound or a complete spectrum is required. Elastic BS-FFT achieves a balance between speed and quality. Density Map is fast and requires minimal development effort but is accurate only \textcolor{black}{at relatively high sampling rates}.

\textcolor{black}{\hypersetup{allcolors=black}Figure~\ref{fig:gnn15-hellinger-time-tradeoff} illustrates the trade-off between estimation quality} and computational cost. Density Map is computationally efficient, but its error increases as the sampling rate decreases. \textcolor{black}{Elastic BS-FFT incurs higher runtime} at the same sampling rate, but achieves comparable accuracy with fewer samples. Together, these two methods provide flexible choices under different time budgets. In contrast, the other methods are either substantially slower or less accurate, making them less competitive across the evaluated scenarios.
\par\endgroup

\section{Related Work}
\begingroup
\emergencystretch=2em

\paragraph{Fourier analysis of 2-D patterns.}
Fourier features have long been used to analyze 2-D patterns. Oliva and Torralba~\cite{Oliva2001SpatialEnvelope} summarize spectral scale and orientation to characterize scene structure, while Manjunath and Ma~\cite{Manjunath1996TextureFeatures} use statistics of Gabor responses at multiple scales and orientations for texture retrieval.
Zhang and Shen~\cite{zhang2026spectralanalysissparsematrix} apply a 2-D FFT to binary sparsity patterns and relate spectral scale, orientation, and entropy to sparse matrix structure. They demonstrate the benefits of adding spectral features to spatial features for SpMV kernel selection on SuiteSparse and pruned LLM matrices. Building on these findings, we investigate the scalability and accuracy of spectral extraction for 2-D sparse inputs, with potential applications not only to sparse matrices but also to compressed images and other sparse data representations.

\paragraph{FFT computation and compact spectra.}
Previous work has developed efficient dense, sparse, nonuniform, and distributed FFT algorithms. Section~\ref{sec:existing-fft-methods} discusses existing sparse, nonuniform, and distributed FFT methods and limitations. There are other sparse FFT methods: sFFT-DT~\cite{Hsieh2014SFFTDT} recovers exactly or approximately sparse one-dimensional spectra, with an analysis that assumes uniformly distributed nonzero frequencies; MIT sFFT~\cite{Hassanieh2012PracticalSFFT} estimates the largest coefficients of a one-dimensional transform. However, these methods do not specifically target 2-D scenarios and therefore lack the necessary optimizations and implementation considerations for such workloads, making them difficult to apply directly in our setting. Therefore, we use FPS-SFT as the coefficient-selecting baseline.

\par\endgroup

\section{Conclusion}

\begingroup
\emergencystretch=2em
In this work, we present the first scalable ladder for FFT-based structural analysis of large sparse binary matrices on modern GPUs. We provide BS-FFT, Elastic BS-FFT, and Density Map for efficient lossless computation, acceleration through spectrum sampling, and spatial compression, respectively. We also provide spectrum reconstruction and direct feature extraction. Experimental results show that these methods together can meet user needs under different time and resource budgets.

\par\endgroup
\clearpage

\bibliographystyle{ACM-Reference-Format}
\bibliography{ref}

\end{document}